\documentstyle[12pt]{article}
\raggedright
\begin{document}

\title{Explicit Spin Coordinates}

\author{\thanks{Author to whom correspondence should be addressed;
email: ghunter@yorku.ca}
Geoffrey Hunter and Ian Schlifer\\
{\small\sl Department of Chemistry},\\
{\small\sl York University, Toronto, Canada  M3J 1P3}}


\maketitle

\vspace*{-0.25in}

\begin{abstract}
The recently established existence of spherical harmonic functions, 
$Y_\ell^{m}(\theta,\phi)$ 
for half-odd-integer values of $\ell$ and $m$, allows for the introduction
into quantum chemistry of explicit electron spin-coordinates; 
i.e.~spherical polar angles 
$\theta_s, \phi_s$, that specify the orientation of the spin angular momentum
vector in space.

In this coordinate representation the spin angular momentum
operators, $S^2, S_z$, are represented by the usual differential operators
in spherical polar coordinates (commonly used for $L^2, L_z$), 
and their electron-spin eigenfunctions are 
$\sqrt{\sin\theta_s} \exp(\pm\phi_s/2)$.

	This eigenfunction representation has the pedagogical advantage over
the abstract spin eigenfunctions, $\alpha, \beta,$ that ``integration over
spin coordinates'' is a true integration (over the angles $\theta_s, \phi_s$).
In addition they facilitate construction of many electron wavefunctions in which
the electron spins are neither parallel nor antiparallel, but inclined 
at an intermediate angle.  In particular this may have application to the
description of EPR correlation experiments.

\end{abstract}

\newpage

\section{The Coordinate Representation of Spin}

The nature of the electron remains an enigma \cite{MacGregor}; however 
(and notwithstanding the Pauli and Dirac representations of spin
\cite{Schiff}) 
the closest physical model is a spherically symmetric distribution of charge 
rotating around an axis \cite[p.55]{Bates}.   This internal rotation is known 
as the electron's spin; its quantized angular momentum component is known to 
be $\pm\hbar/2$ (associated with a magnetic moment $\hbar e/2mc$).

In the coordinate representation of spin, the direction of the electron's 
axis of rotation may be specified by two spherical polar angles: 
$\theta_s, \phi_s$; the subscript, $s$, distinguishes these angles from 
the $\theta, \phi$ that describe the angular position of the electron relative
to the nucleus in the Schr\"odinger theory of the hydrogen atom;
the $\theta, \phi$ coordinate system has its origin at the nucleus,
whereas the $\theta_s, \phi_s$ coordinate system has its origin at the 
electron.  

The recent discovery of spherical harmonics, 
$Y_\ell^{m}(\theta,\phi)$, 
for half-odd-integer values of $\ell$ and $m$ \cite{JPhysA}, provides the
basis for a coordinate representation of electron spin, the coordinates
being  the two spherical polar angles $\theta_s, \phi_s$.

In this representation the operators for the square of the total spin 
angular momentum, $\:{\bf S^2}$, and its $z$-component, $\:{\bf S_z}$, are
expressed in terms of the spherical polar angles \cite[p.207]{McQuarrie}, 
\cite[p.95]{Rae} by:
\begin{eqnarray}\label{amops} 
{\bf S^2} = - \hbar^2
\left\{
\frac{1}{\sin \theta_s }\frac{\partial}{\partial\theta_s}\left(\sin \theta_s 
\frac{\partial}{\partial\theta_s}\right) +
\frac{1}{\sin^2 \theta_s }\frac{\partial^2}{\partial\phi_s^2}
\right\} \quad\quad
{\bf S_z} = \frac{\hbar}{\rm i}\frac{\partial}{\partial\phi_s} 
\end{eqnarray}
The eigenfunctions of these operators, 
$\alpha(\theta_s,\phi_s)$, $\beta(\theta_s,\phi_s)$,
corresponding to the known spin-states of an electron are \ref{JPhysA}:
\begin{eqnarray}\label{fsh}
\alpha(\theta_s,\phi_s)=\frac{1}{\pi}\sqrt{\sin\theta_s} e^{i(+\phi_s/2)}
\quad\quad{\rm for}\ \ m_s=+1/2\\\nonumber
\beta(\theta_s,\phi_s)=\frac{1}{\pi}\sqrt{\sin\theta_s} e^{i(-\phi_s/2)}
\quad\quad{\rm for}\ \ m_s=-1/2
\end{eqnarray}
where $m_s$ is the well-known spin quantum number, and the factor of $1/\pi$
normalizes the functions with respect to integration over $\theta_s$ and 
$\phi_s$ (see below).

Application of the operators (\ref{amops}) to the functions (\ref{fsh}) 
shows that both of these functions are eigenfunctions:
\begin{itemize}
\item of $S^2$ with eigenvalue $3\hbar^2/4$, and 
\item of $S_z$ with eigenvalues of $m_s\hbar = \pm\hbar/2$.
\end{itemize}
The functions (\ref{fsh}) are normalized in the sense that:
\begin{equation}\label{normcheck}
\int_0^{2\pi}\,d\phi_s \int_0^{\pi} 
Y^*(\theta_s,\phi_s)\: Y(\theta_s,\phi_s)\: \sin(\theta_s)\,d\theta_s = 1
\end{equation}
in which $^*$ denotes the complex conjugate, and $Y(\theta_s,\phi_s)$ is
one of $\alpha(\theta_s,\phi_s)$ or $\beta(\theta_s,\phi_s)$.
They are also orthogonal in the sense that:
\begin{equation}\label{orthogonal}
\int_0^{2\pi}
\alpha^*(\theta_s,\phi_s)\: \beta(\theta_s,\phi_s) 
\,d\phi_s  
= 
\int_0^{2\pi}
\alpha(\theta_s,\phi_s)\: \beta^*(\theta_s,\phi_s) 
\,d\phi_s  
= 0
\end{equation}

The normalization (\ref{normcheck}) is the one usually used for the 
spherical harmonics.
However, it should be noted that the Fermion functions (\ref{fsh}) are only
single-valued when $\phi$ is considered to have the range of a double circle:  
$0$$\le$$\phi_s$$\le$$4\pi$; this $\:4\pi\:$ symmetry is a well known property
of fermion wavefunctions \cite[p.21 \& p.138]{Icke}.
Alternatively, if the range of $\phi_s$ is regarded as that of a single circle,
$0$$\le$$\phi_s$$\le$$2\pi$, they are double-valued functions in the sense
that:
\begin{eqnarray}\nonumber
Y(\theta_s,\phi_s+2\pi)=-Y(\theta_s,\phi_s)
\end{eqnarray}
The double-valuedness
is admissible because the probability density, \\\noindent
$Y^*(\theta_s,\phi_s)\,Y(\theta_s,\phi_s)$ is single-valued and positive.
The alternative of basing the normalization upon 
$0$$\le$$\phi_s$$\le$$4\pi$
would simply introduce a factor of $\sqrt{2}$ into the normalization constant.

\section{Previous Representation of Spin}

The history of intrinsic (spin) angular momentum is a fascinating story 
\cite{Tomonaga}.  
  The spin of an electron has been represented most 
simply by eigenfunctions, $\:\alpha\:$ and $\:\beta$, which are abstract 
in the sense 
that they are not functions of any coordinates; they are defined to be
eigenfunctions of $\:{\bf S^2}\:$ and $\:{\bf S_z}\:$ with eigenvalues 
of $\:\hbar^2 s(s+1)\:$
and $\:\hbar m_s$, with $\:s=\frac{1}{2}\:$ and $\:m_s=\pm\frac{1}{2}\:$
\cite[300]{McQuarrie}:
\begin{eqnarray}\label{absamops} 
{\bf S^2} 
\left\{\begin{array}{c}\alpha\\\beta\end{array}\right\} 
= \hbar^2 {\textstyle\frac{1}{2}}({\textstyle\frac{1}{2}}+1)
\left\{\begin{array}{c}\alpha\\\beta\end{array}\right\} 
\quad\quad
{\bf S_z} 
\left\{\begin{array}{c}\alpha\\\beta\end{array}\right\} 
= 
\left\{\begin{array}{c} + \\ - \end{array}\right\} 
{\textstyle\frac{\displaystyle\hbar}{2}}
\left\{\begin{array}{c}\alpha\\\beta\end{array}\right\} 
\end{eqnarray}
Different authors handle the abstract nature of the spin eigenfunctions,
$\alpha$ and $\beta$, differently.  In presenting the orthonormality
relations:
\begin{eqnarray}\label{norm} 
<\alpha|\alpha> = <\beta|\beta> = 1\\\label{ortho} 
<\alpha|\beta> = <\beta|\alpha> = 0
\end{eqnarray}
Dykstra \cite[p.185]{Dykstra} avoids explicit consideration of
the functional dependence of the ``abstract functions'' $\alpha$ and $\beta$ 
on the ``abstract spin coordinate'', and likewise avoids explicit 
consideration of the integration ``over the spin coordinate'' implicit
in (\ref{norm}) and (\ref{ortho}).

In this purely abstract interpretation of 
$\alpha$ and $\beta$ the simplest concept is to define them by the eigenvalue
equations (\ref{absamops}) with normalization defined by (\ref{norm}) 
(their orthogonality, (\ref{orthogonal}), follows from (\ref{absamops}) 
because they are 
eigenfunctions of $S_z$ with different eigenvalues).  

Expectation
values such as 
$<\alpha|{\bf S^2}|\alpha>$, 
$<\beta|{\bf S^2}|\beta>$, 
$<\alpha|{S_z}|\alpha>$, and
$<\beta|{S_z}|\beta>$, and
(\ref{norm}) are likewise abstract - they do not have to refer 
to ``integration over the spin coordinate''.

Other authors \cite[p.15]{Calais}, \cite[p.14]{Kaempffer}, 
\cite[p.253]{Kramers}, \cite[p.233]{Schiff},
invoke the Pauli 2$\times$2 matrix representation
of electron spin 
to infer that the spin coordinate/variable spans a discrete space of just
2 values, which are $\{1,0\}$ for $\alpha$ and 
$\{0,1\}$ (or $\{0,-1\}$ \cite{Calais}) for $\beta$.  
Calais then correctly says that:
\begin{quote}
``integration over spin space \ldots is a misnomer for summation over spin
space''
\end{quote}
In the Pauli matrix representation this summation 
(in (\ref{norm},\ref{ortho})) is the scalar product of the transpose of 
$\alpha, \beta$ ($\{1,0\}$, $\{0,1\}$) with the 2$\times$1 column
vectors.  Identifying the 2 elements of the column-vector eigenfunctions of the 
Pauli representation as a space of 2 points spanned by the ``spin
coordinate'', is a dubious concept, for as Schiff shows 
\cite[pp.144-147]{Schiff} the number of points of the spin-space increases
with the total angular momentum quantum number, $j$: it is $2\,j+1$ for 
$j=1/2, 1, 3/2$, etc.

In many electron wavefunctions one must designate which electron is in
a spin eigenstate, $\alpha, \beta$.  This is commonly indicated by
$\alpha(s_1), \beta(s_2)$ etc, or simply $\alpha(1), \beta(2)$ etc.
This notation indicates that $\alpha(s_1)$ denotes an $\alpha$ spin function 
for electron $1$, and some authors (naturally) refer to $s_1, s_2$, etc,
as the ``spin coordinate'' of electron 1, 2, etc.  In this context these
electron designators are really specifying labels rather than coordinates.
Similar designators will be needed in our spin-coordinate representation 
of {\em many-electron}
wavefunctions; different angles, $\theta_{s_j}$, $\phi_{s_j}$, 
($j=1, 2, \ldots$) must be used for each electron.

\section{Discussion}

Heretofore no scalar representation in terms of the explicit (angular) 
coordinates of the spin motion has been used, and yet such positional 
coordinates are believed to have a privileged role in the 
interpretation of quantum mechanics \cite{Goldstein}.  

The logical steps in the derivation of the Fermion Spherical Harmonics
\cite{JPhysA} parallel the arguments used to derive the abstract
eigenfunctions and eigenvalues of angular momentum, which (as is well-known)
lead to both integer and half-odd-integer values of $\ell$ and $m$
\cite[\S5.4,p.115]{Levine}.
It was the belief
that the wavefunction (rather than the probability) must be single-valued
that inhibited discovery of these functions for so many years.

We hope that the  explicit coordinates, $\theta_s$, $\phi_s$,
for the orientation of the spin vector of each electron will be adopted 
in the teaching of quantum chemistry, for they will make ``integration
over spin coordinates'' a true integration entirely comparable with
integration over the positional coordinates of each electron (such as
$r,\,\theta,\,\phi$ for an atomic electron), and the spin eigenfunctions
$\alpha$, $\beta$ will have an explicit functional dependence upon the
angles $\theta_s, \phi_s$ as shown explicitly in (\ref{fsh}).

Apart from their pedagogical value, these angular dependent spin
eigenfunctions will facilitate the construction of wavefunctions in
which the orientation of a particular spin vector is neither parallel
to, nor anti-parallel to, another spin vector or to an external magnetic
field.  

One possible application is to the theory of Nuclear Magnetic Resonance (NMR),
where the spins of different nuclei will, in general,
be neither parallel nor anti-parallel.

Another example is the Einstein-Podolski-Rosen gedanken experiment, in which
two electrons go away from a source in opposite directions towards
spin-orientation detectors which are inclined at any angle, $\theta$
(when parallel, $\theta$$=$$0$, or anti-parallel, $\theta$$=$$\pi$).
The experiment measures correlations between detections at the two detectors
as a function of their angle of inclination, $\theta$; further detail is 
beyond the scope of this article \cite[Ch.4,p.35]{Rodrigues}.

\section*{Acknowledgements}

This work was supported by the Natural Sciences and Engineering Research
Council of Canada.


\end{document}